\documentclass[]{aa}
\usepackage{natbib}
\usepackage{graphicx}
\usepackage[a4paper]{hyperref}
\usepackage{txfonts}
\begin{document}
\def\teff{$T\rm_{eff }$}
\def\kms{$\mathrm {km s}^{-1}$}

\title{
The Sgr dSph hosts a metal-rich population
\thanks{Based on observations obtained with UVES
at VLT Kueyen 8.2m telescope in program 67.B-0147}\fnmsep
\thanks{Table 2 is also available in electronic form 
at the CDS via anonymous ftp to {\tt cdsarc.u-strasbg.fr} 
(130.79.128.5) or via  
\href{http://cdsweb.u-strasbg.fr/cgi-bin/qcat?J/A+A/}{http://cdsweb.u-strasbg.fr/cgi-bin/qcat?J/A+A/}
}}

   \subtitle{}

\author{
P. \,Bonifacio \inst{1},
L. \,Sbordone  \inst{2,3,4}
G. \,Marconi  \inst{4}
L. \,Pasquini  \inst{4}
\and V. Hill\inst{5}
          }

  \offprints{P. Bonifacio}

\institute{
Istituto Nazionale di Astrofisica --
Osservatorio Astronomico di Trieste, Via Tiepolo 11,
I-34131 Trieste, Italy
\and
Istituto Nazionale di Astrofisica --
Osservatorio Astronomico di Roma, Italy
Via Frascati 33, 00040 Monteporzio Catone, Roma
\and
Universit\`a Tor Vergata, Roma
\and
European Southern Observatory, Casilla 19001, Santiago, Chile
\and
Observatoire de Paris- Meudon,  2 pl. J. Janssen, 92190, France
}

\authorrunning{Bonifacio et al.}
\mail{bonifaci@ts.astro.it}

\titlerunning{Sgr dSph}

\date{Received  / Accepted }

\abstract{  We report on abundances of  O, Mg, Si, Ca and Fe for
10 giants in the Sgr dwarf spheroidal derived from high resolution
spectra obtained with UVES
at the  8.2m Kueyen-VLT telescope.
The iron abundance spans the range $\rm -0.8 \la [Fe/H] \la 0.0$
and the dominant population  is relatively
metal-rich with [Fe/H]$\sim -0.25$.
The $\alpha$/Fe ratios are slightly subsolar, even at the lowest
observed metallicities suggesting a slow or bursting star formation rate.
From our sample of 12 giants (including the two observed by \citealt{B00})
we conclude that a substantial metal rich
population exists in Sgr, which dominates the sample. 
The spectroscopic metallicities allow  one to break the age-metallicity
degeneracy in the interpretation of the colour-magnitude diagram (CMD).
Comparison of isochrones of appropriate metallicity with the observed
CMD suggests an age of 1 Gyr or younger, for the dominant  Sgr population
sampled by us.
We argue that the observations support a star formation that is triggered
by the passage of Sgr through the Galactic disc, both in Sgr and in the
disc. This scenario has also the virtue of explaining the  mysterious
``bulge C stars'' as disc stars formed in this event. 
The interaction 
of Sgr with the Milky Way is likely to have played a major role
in its evolution.
\keywords{Stars: abundances --
Stars: atmospheres -- Galaxies: abundances -- Galaxies: evolution --
Galaxies: dwarf -- Galaxies:individual Sgr dSph}
}
\maketitle{}

\begin{table}
\caption{Log of observations (only for the stars not described in \citet{B00})}
\label{logobs}
\begin{center}
{\scriptsize
\begin{tabular}{lrrrr}
\hline
\\
Star & date & hour & exp.time (secs) & r. v. (km/s) \\
\\
\hline
\\
432      & 2001 - 06 - 28 & 02:08:43 & 3800    & 160.6 \\ 
         &                & 03:15:25 & 3600    & 160.6 \\
	 & 2001 - 07 - 24 & 04:17:08 & 3600    & 161.1 \\
\\
628      & 2001 - 07 - 19 & 04:02:17 & 3600    & 146.2 \\ 
         &                & 05:03:35 & 3600    & 145.7 \\
	 & 2001 - 07 - 24 & 00:25:25 & 3600    & 145.1 \\
\\
635      & 2001 - 07 - 19 & 00:47:21 & 3600    & 125.1 \\ 
         &                & 01:49:14 & 3600    & 125.4 \\
	 &                & 02:51:14 & 3600    & 125.1 \\
\\
656      & 2001 - 06 - 29 & 01:46:34 & 3600    & 134.9 \\ 
         &                & 02:49:10 & 3600    & 134.7 \\
	 &                & 03:55:24 & 3600    & 134.8 \\
	 & 2001 - 07 - 18 & 02:21:57 & 3600    & 135.4 \\
\\
709      & 2001 - 06 - 29 & 05:01:13 & 3600    & 124.5 \\ 
         & 2001 - 07 - 17 & 02:04:06 & 3600    & 125.7 \\
	 &                & 03:09:44 & 3600    & 125.4 \\
	 & 2001 - 07 - 18 & 01:15:04 & 3600    & 125.4 \\
\\
716      & 2001 - 07 - 21 & 01:06:59 & 3600    & 135.2 \\     
         & 2001 - 07 - 22 & 01:29:06 & 3600    & 135.6 \\
	 &                & 02:35:01 & 3600    & 136.0 \\
\\
717      & 2001 - 07 - 24 & 01:31:00 & 3600    & 141.9 \\     
         &                & 02:33:38 & 3600    & 141.7 \\
	 &                & 03:39:32 & 3600    & 142.4 \\
\\
867      & 2001 - 07 - 18 & 03:30:28 & 3600    & 154.5 \\     
         &                & 04:35:41 & 3600    & 154.2 \\
	 &                & 05:38:57 & 3600    & 154.1 \\
\\
894      & 2001 - 06 - 26 & 02:14:30 & 3600    & 128.4 \\     
         & 2001 - 06 - 27 & 04:26:59 & 3600    & 129.6 \\
	 &                & 07:05:23 & 3600    & 130.2 \\
	 & 2001 - 07 - 17 & 06:05:14 & 3600    & 129.8 \\
\\
927      & 2001 - 07 - 14 & 03:09:11 & 3600    & 144.0 \\     
         & 2001 - 07 - 15 & 02:32:59 & 3600    & 144.2 \\
	 &                & 03:38:19 & 3600    & 143.6 \\
	 & 2001 - 07 - 17 & 00:58:34 & 3600    & 144.3 \\
\\
\hline

\end{tabular}
}
\end{center}
\end{table}

\section{Introduction}

The Sagittarius dwarf spheroidal galaxy is the nearest satellite
of the Milky Way. Right from  its discovery \citep{ibata94,ibata95} 
it was apparent that it displayed a wide red giant branch,
which has been  interpreted as evidence for a dispersion in metallicity.
The chemical composition  and, in particular, abundance ratios,
of a stellar population contains important information
on its star formation history and evolution.
The RGB of Sgr is within reach of the high resolution spectrographs
operating at 8m class telescopes. \citet{B00} reported the first
detailed chemical abundances for two Sgr giants using the data
taken during the commissioning of UVES\citep{dekker}. 
Contrary to our expectations
the two stars turned out to be very similar and of relatively
high metallicity ([Fe/H]$\sim -0.25$), considerably more metal-rich
than the highest photometric metallicity. The other striking result
was that both stars showed a low value of the $\alpha$ elements to iron
ratios. Clearly no general conclusions may be drawn from a sample
of two stars. For this reason we undertook a program  to observe
other Sgr giants  at high resolution with UVES. 

In this paper we report  on the abundances of O, Mg, Si, Ca and Fe for
10 giants which are confirmed radial velocity members of Sgr
\citep{B99,bonivlt}.

\section{Observations}

\begin{figure*}
\centering
\resizebox{\hsize}{!}{\includegraphics[clip=true]{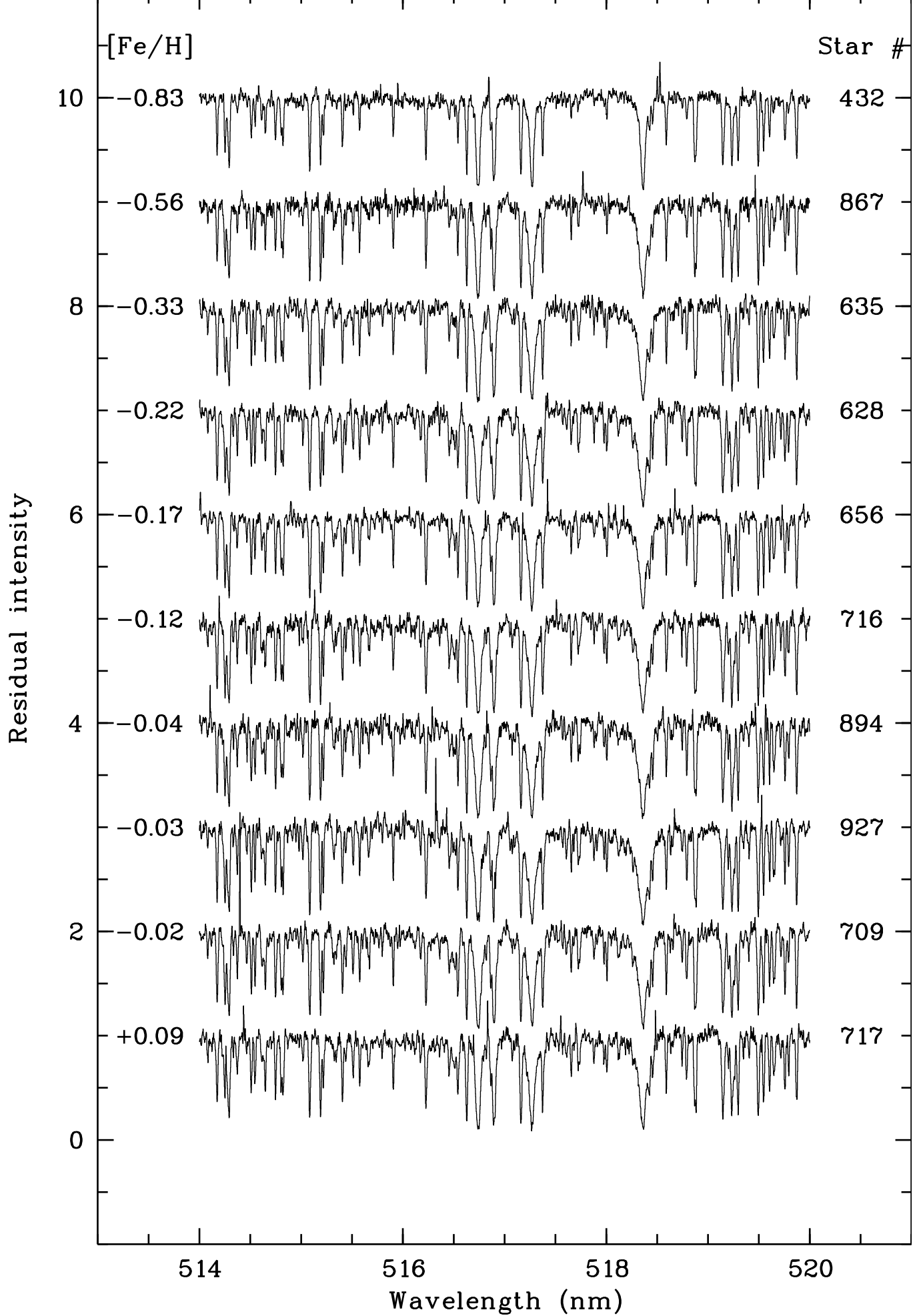}}
\caption{Coadded UVES spectra of the 10 Sgr giants analysed in this 
paper. The labels on the right denote the star number, those on the
left the [FeI/H].}
\label{data}
\end{figure*}

The data were obtained between 25 June and 23 July 2001, 
under good seeing conditions, in service mode as detailed in
Table \ref{logobs}. 
Both blue and red arm spectra were acquired
(dichroic mode DIC1)
but only the red arm data have been analyzed so far. 
We used the standard setting with central wavelength at 580 nm, 
covering the range between 480 and 680 nm, and providing a resolution
of about 43000  with a slit width of 1''. 
A $2\times 2$ 
on-chip binning has been used, without loss in resolution (due to the
rather wide slit).
For each star, three or four one-hour exposure have been taken. 
The UVES red arm detector is a mosaic of two $4096\times2048$ CCDs, 
covering (in the setting we used) the spectral ranges
480-580 and 580-680 nm. 
The region of the Mg I b triplet is shown for all the observed stars,
ordered by metallicity,  in Fig. \ref{data}.

We also analysed FEROS spectra of two Hyades stars,  HD 27371 and HD 27697,
to provide a reference for our metallicity scale. The spectra
were acquired at the ESO 1.5m telescope on October 30th 2000, with
120s exposure. They 
have a resolution R$\sim$ 48000 and cover the range 370 nm -- 910 nm,
however we used only the range 530 nm - 700 nm which provides 
a S/N ratio of about 150.

\begin{table*}
\begin{center}
\caption{Lines}
\label{abund1}
{\scriptsize
\begin{tabular}{rrrlrrrrrrrrrr}
\hline
\\
Ion  & $\lambda$ & log gf  & source of    & EW      & $\epsilon$ & EW       & $\epsilon$ & EW      & $\epsilon$ & EW      & $\epsilon$ & EW     & $\epsilon$ \\
     & (nm)      &         & log gf       & (pm)    &            & (pm)     &            & (pm)    &            & (pm)    &            & (pm)   &           \\
     &           &         & (see notes)  & 432     &            & 628      &            & 635     &            & 656     &            & 709    &           \\
\\
\hline
\\
Fe I & 489.2871  & -1.29   & FMW          & 4.42    & 6.69       & 8.28    & 7.34      & --      & --         & 7.78    & 7.44       & 7.84    & 7.47       \\
Fe I & 506.7151  & -0.97   & FMW          & --      & --         & 10.51   & 7.41      & 8.42    & 7.03       & 9.54    & 7.46       & 9.52    & 7.49       \\
Fe I & 510.4436  & -1.69   & FMW          & 2.46    & 6.72       & --      & --        & 4.51    & 7.14       & 4.81    & 7.35       & 6.28    & 7.62       \\
Fe I & 510.9650  & -0.98   & FMW          & 6.07    & 6.82       & --      & --        & 9.51    & 7.34       & 7.98    & 7.24       & 10.0    & 7.69       \\
Fe I & 552.5539  & -1.33   & FMW          & 3.51    & 6.52       & 7.17    & 7.16      & 8.14    & 7.30       & 7.92    & 7.45       & 7.43    & 7.37       \\
Fe I & 585.5091  & -1.76   & FMW          & --      & --         & --      & --        & --      & --         & --      & --         & 5.02    & 7.76       \\
Fe I & 585.6083  & -1.64   & FMW          & 2.38    & 6.63       & 6.16    & 7.38      & 5.52    & 7.23       & 4.85    & 7.28       & 6.61    & 7.57       \\
Fe I & 585.8779  & -2.26   & FMW          & --      & --         & 2.56    & 7.30      & 2.88    & 7.30       & 2.00    & 7.24       & 1.78    & 7.14       \\
Fe I & 586.1107  & -2.45   & FMW          & --      & --         & --      & --        & --      & --         & --      & --         & 2.84    & 7.66       \\
Fe I & 587.7794  & -2.23   & FMW          & 1.65    & 6.88       & --      & --        & --      & --         & 3.64    & 7.53       & 4.44    & 7.64       \\
Fe I & 588.3813  & -1.36   & FMW          & 5.80    & 6.70       & 10.41   & 7.38      & --      & --         & 8.48    & 7.26       & 8.85    & 7.35       \\
Fe I & 615.1617  & -3.23   & FMW          & 6.81    & 6.66       & 9.92    & 7.04      & --      & --         & 9.37    & 7.22       & 8.66    & 7.06       \\
Fe I & 616.5361  & -1.55   & FMW          & 4.19    & 6.75       & --      & --        & 6.22    & 7.06       & 6.39    & 7.27       & 6.64    & 7.30       \\
Fe I & 618.7987  & -1.72   & FMW          & 4.01    & 6.65       & --      & --        & 7.51    & 7.20       & 6.49    & 7.22       & 8.42    & 7.56       \\
Fe I & 649.6469  & -0.57   & FMW          & 3.56    & 6.39       & 9.10    & 7.29      & 7.73    & 7.07       & 7.81    & 7.24       & 9.40    & 7.54       \\
Fe I & 670.3568  & -3.16   & FMW          & 4.09    & 6.67       & 7.53    & 7.23      & 6.29    & 7.00       & 7.27    & 7.37       & --      & --         \\
\\
Fe II & 483.3197  & -4.78   & FMW          & 1.20    & 6.76       & --      & --        & --      & --         & 3.18    & 7.49       & 3.50    & 7.61       \\
Fe II & 492.3927  & -1.32   & FMW          & 14.99   & 6.50       & --      & --        & --      & --         & --      & --         & --      & --         \\
Fe II & 499.3358  & -3.65   & FMW          & 4.10    & 6.61       & 8.35    & 7.50      & 6.35    & 7.04       & 7.05    & 7.30       & 7.35    & 7.48       \\
Fe II & 510.0664  & -4.37   & FMW          & 2.08    & 6.82       & 3.76    & 7.38      & 4.30    & 7.41       & 3.30    & 7.26       & 3.41    & 7.34       \\
Fe II & 513.2669  & -4.18   & FMW          & 2.57    & 6.76       & 4.43    & 7.32      & 4.01    & 7.16       & 4.72    & 7.36       & 3.81    & 7.24       \\
Fe II & 516.1184  & -4.48   & K88          & --      & --         & --      & --        & --      & --         & 2.17    & 7.15       & --      & --         \\
Fe II & 525.6938  & -4.25   & K88          & 2.57    & 6.93       & 5.44    & 7.67      & 3.36    & 7.19       & 4.66    & 7.51       & 3.73    & 7.39       \\
Fe II & 526.4812  & -3.19   & FMW          & 4.84    & 6.79       & 7.10    & 7.28      & --      & --         & --      & --         & 7.16    & 7.43       \\
Fe II & 614.9258  & -2.72   & K88          & 3.69    & 6.81       & --      & --        & 3.55    & 6.81       & --      & --         & 5.67    & 7.38       \\
\\
Mg I & 552.8405  & -0.620  & LZ            & --      & --         & 21.74   & 7.09      & 22.55   & 7.07       & 19.77   & 7.09       & 21.53   & 7.20       \\
Mg I & 571.1088  & -1.833  & LZ            & 7.98    & 6.75       & 12.46   & 7.29      & 10.81   & 7.06       & 11.05   & 7.27       & 10.66   & 7.20       \\
Mg I & 631.8717  & -1.981  & FF            & 2.08    & 6.73       & 5.04    & 7.32      & 5.83    & 7.40       & 3.59    & 7.16       & 5.95    & 7.51       \\
Mg I & 631.9237  & -2.201  & FF            & --      & --         & 4.77    & 7.50      & 3.66    & 7.28       & --      & --         & 3.61    & 7.35       \\
\\
Si I & 577.2146  & -1.750  & GARZ          & --      & --         & --      & --        & --      & --         & --      & --         & 4.82    & 7.22       \\
Si I & 594.8541  & -1.230  & GARZ          & 6.16    & 6.79       & 8.42    & 7.18      & 8.26    & 7.12       & 9.30    & 7.38       & 8.84    & 7.35       \\
Si I & 612.5021  & -1.540  & ED            & 1.80    & 6.83       & --      & --        & --      & --         & --      & --         & --      & --         \\
Si I & 614.2483  & -1.480  & ED            & --      & --         & 4.54    & 7.45      & 3.21    & 7.17       & 3.50    & 7.28       & 4.34    & 7.46       \\
Si I & 614.5016  & -1.430  & ED            & 1.39    & 6.59       & --      & --        & 3.28    & 7.13       & 3.34    & 7.20       & 3.79    & 7.30       \\
Si I & 615.5134  & -0.770  & ED            & 3.27    & 6.42       & 8.26    & 7.30      & 5.64    & 6.88       & 7.99    & 7.32       & 7.87    & 7.34       \\
\\
Ca I & 551.2980  & -0.290  & NBS            & 5.48    & 5.06       & 10.74   & 5.79      & 10.64   & 5.80       & 10.69   & 6.01       & 10.47   & 5.96       \\
Ca I & 585.7451  & 0.230   & NBS            & --      & --         & --      & --        & 14.22   & 5.86       & --      & --         & 17.63   & 6.51       \\
Ca I & 586.7562  & -1.610  & ED             & --      & --         & 3.86    & 6.08      & 3.22    & 5.92       & 4.26    & 6.26       & 4.68    & 6.28       \\
Ca I & 612.2217  & -0.409  & NBS            & --      & --         & --      & --        & 19.53   & 5.95       & --      & --         & 19.51   & 6.25       \\
Ca I & 616.1297  & -1.020  & NBS            & 5.98    & 5.38       & --      & --        & 9.33    & 5.81       & 9.98    & 6.12       & 10.46   & 6.19       \\
Ca I & 616.6439  & -0.900  & NBS            & 6.75    & 5.40       & 10.81   & 5.89      & 9.54    & 5.72       & 9.73    & 5.96       & 10.29   & 6.03       \\
Ca I & 616.9042  & -0.550  & NBS            & 7.25    & 5.15       & 13.05   & 5.88      & 10.88   & 5.60       & 11.64   & 5.94       & 12.11   & 6.01       \\
Ca I & 643.9075  & 0.470   & NBS            & --      & --         & --      & --        & 17.94   & 5.67       & --      & --         & 18.91   & 6.07       \\
Ca I & 645.5558  & -1.350  & NBS            & --      & --         & 8.98    & 6.05      & 8.05    & 5.92       & 8.16    & 6.13       & 8.02    & 6.07       \\
Ca I & 649.3781  & 0.140   & NBS            & --      & --         & --      & --        & 15.03   & 5.57       & --      & --         & --      & --         \\
Ca I & 649.9650  & -0.590  & NBS            & 7.69    & 5.27       & --      & --        & 10.91   & 5.62       & 10.92   & 5.82       & 12.82   & 6.15       \\
Ca I & 650.8850  & -2.110  & NBS            & 1.38    & 5.46       & 2.48    & 5.83      & --      & --         & 2.01    & 5.82       & 2.31    & 5.83       \\
Ca I & 679.8479  & -2.320  & K88            & --      & --         & --      & --        & --      & --         & 1.45    & 6.05       & --      & --         \\
\\
\hline
\\
\multispan{10}{ED    Edvardsson, B. et al. 1993, A. and A. 275, 101 \hfill}\\
\multispan{10}{FF    Froese Fischer, C. 1975, Can.J.Phys. 53, 184; p. 338; 1979, JOSA 69, 118. \hfill}\\
\multispan{10}{FMW   Fuhr, J.R., Martin, G.A., and Wiese, W.L. 1988. J.Phys.Chem.Ref.Data 17, Suppl. 4. \hfill}\\
\multispan{10}{GARZ  Garz, T. 1973, a. and A. 26, 471. \hfill}\\
\multispan{10}{K88   Kurucz, R.L. 1988, Trans. IAU, XXB, M. McNally, ed., Dordrecht: Kluwer, 168-172. \hfill}\\
\multispan{10}{KZ    Kwiatkowski, M., Zimmermann, P., Biemont, E., and Grevesse, N. 1982, A. and A. 112, 337-340.}\\
\multispan{10}{LZ    Lincke, R. and Ziegenbein, G. 1971, Z. Phyzik, 241, 369. \hfill}\\
\multispan{10}{NBS   Wiese, W.L., Smith, M.W., and Glennon, B.M. 1966, NSRDS-NBS 4. \hfill}\\
\end{tabular}
}
\end{center}
\end{table*}

\addtocounter{table}{-1}

\begin{table*}
\begin{center}
\caption{Lines (continued)}
\label{abund1bis}
{\scriptsize
\begin{tabular}{rrrlrrrrrrrrrr}
\hline
\\
Ion  & $\lambda$ & log gf  & source of   & EW      & $\epsilon$ & EW       & $\epsilon$ & EW (pm) & $\epsilon$ & EW      & $\epsilon$ & EW      & $\epsilon$ \\
     & (nm)      &         & log gf      & (pm)    &            & (pm)     &            & (pm)    &            & (pm)    &            & (pm)    &            \\\
     &           &         & (see notes) & 716     &            & 717      &            & 867     &            & 894     &            & 927    &            \\
\\
\hline
\\
Fe I & 489.2871  & -1.29   & FMW          & 8.67    & 7.45       & 7.21    & 7.50      & 6.89    & 7.14       & 8.21    & 7.57       & 7.10    & 7.44       \\
Fe I & 506.7151  & -0.97   & FMW          & 11.31   & 7.59       & --      & --        & --      & --         & 8.82    & 7.37       & 9.00    & 7.56       \\
Fe I & 510.4436  & -1.69   & FMW          & --      & --         & 6.32    & 7.75      & 5.12    & 7.23       & --      & --         & 5.90    & 7.62       \\
Fe I & 510.9650  & -0.98   & FMW          & 8.58    & 7.20       & 8.94    & 7.67      & 8.06    & 7.02       & 9.39    & 7.59       & 8.38    & 7.50       \\
Fe I & 552.5539  & -1.33   & FMW          & 8.83    & 7.46       & 7.41    & 7.52      & 6.75    & 7.04       & 7.06    & 7.30 .     & 6.92    & 7.38       \\
Fe I & 585.5091  & -1.76   & FMW          & --      & --         & --      & --        & --      & --         & 4.65    & 7.69       & 4.09    & 7.63       \\
Fe I & 585.6083  & -1.64   & FMW          & --      & --         & 5.86    & 7.57      & 5.15    & 7.16       & 6.89    & 7.64       & 6.50    & 7.66       \\
Fe I & 585.8779  & -2.26   & FMW          & 2.89    & 7.41       & 2.97    & 7.52      & --      & --         & 2.68    & 7.34       & 1.91    & 7.18       \\
Fe I & 586.1107  & -2.45   & FMW          & 1.60    & 7.35       & 2.85    & 7.75      & --      & --         & 2.47    & 7.56       & 2.49    & 7.60       \\
Fe I & 587.7794  & -2.23   & FMW          & 3.73    & 7.48       & --      & --        & 1.77    & 6.96       & --      & --         & 3.94    & 7.58       \\
Fe I & 588.3813  & -1.36   & FMW          & 10.38   & 7.42       & 9.18    & 7.63      & 6.13    & 6.65       & 9.33    & 7.47       & --      & --         \\
Fe I & 615.1617  & -3.23   & FMW          & 10.86   & 7.26       & --      & --        & 8.04    & 6.73       & --      & --         & 9.05    & 7.27       \\
Fe I & 616.5361  & -1.55   & FMW          & --      & --         & 7.64    & 7.65      & 4.27    & 6.75       & 6.58    & 7.28       & 6.14    & 7.29       \\
Fe I & 618.7987  & -1.72   & FMW          & 8.21    & 7.38       & 6.80    & 7.41      & 4.92    & 6.79       & 6.74    & 7.25       & 6.69    & 7.33       \\
Fe I & 649.6469  & -0.57   & FMW          & 8.50    & 7.23       & 8.07    & 7.46      & 5.75    & 6.73       & 9.06    & 7.51       & 7.66    & 7.35       \\
Fe I & 670.3568  & -3.16   & FMW          & 7.80    & 7.32       & 7.83    & 7.66      & 7.00    & 7.10       & 7.61    & 7.37       & 8.36    & 7.64       \\
\\
Fe II & 499.3358  & -3.65  & FMW          & 7.14    & 7.25       & 6.32    & 7.27      & 6.96    & 6.99       & 5.78    & 7.21       & 5.47    & 7.33       \\
Fe II & 510.0664  & -4.37  & FMW          & 4.64    & 7.51       & 4.13    & 7.49      & 3.63    & 7.11       & 4.34    & 7.60       & --      & --         \\
Fe II & 513.2669  & -4.18  & FMW          & 4.64    & 7.32       & 4.16    & 7.30      & 4.66    & 7.11       & 4.67    & 7.49       & 3.34    & 7.34       \\
Fe II & 516.1184  & -4.48  & K88          & --      & --         & --      & --        & --      & --         & 2.46    & 7.31       & 2.12    & 7.35       \\
Fe II & 525.6938  & -4.25  & K88          & --      & --         & 5.06    & 7.67      & --      & --         & 2.93    & 7.24       & 3.20    & 7.46       \\
Fe II & 526.4812  & -3.19  & FMW          & --      & --         & 7.55    & 7.53      & 6.11    & 6.85       & 6.32    & 7.33       & 5.27    & 7.28       \\
Fe II & 614.9258  & -2.72  & K88          & 5.80    & 7.25       & 7.29    & 7.70      & --      & --         & 4.80    & 7.26       & --      & --         \\
\\
Mg I & 552.8405  & -0.620  & LZ            & 21.90   & 7.14       & --      & --        & 17.65   & 6.71       & 19.47   & 7.03       & 19.47   & 7.02       \\
Mg I & 571.1088  & -1.833  & LZ            & 11.76   & 7.24       & 12.42   & 7.59      & 10.73   & 7.05       & 10.49   & 7.16       & 11.73   & 7.40       \\
Mg I & 631.8717  & -1.981  & FF            & 4.53    & 7.27       & 5.34    & 7.49      & 2.40    & 6.83       & 5.32    & 7.40       & 4.97    & 7.37       \\
Mg I & 631.9237  & -2.201  & FF            & --      & --         & 3.78    & 7.44      & 2.28    & 7.02       & --      & --         & --      & --         \\
\\
Si I & 577.2146  & -1.750  & GARZ          & --      & --         & --      & --        & --      & --         & --      & --         & 4.82    & 7.30       \\
Si I & 594.8541  & -1.230  & GARZ          & 10.26   & 7.44       & 10.14   & 7.62      & 6.71    & 6.83       & 9.24    & 7.45       & 8.98    & 7.49       \\
Si I & 612.5021  & -1.540  & ED            & --      & --         & --      & --        & 4.70    & 6.82       & --      & --         & --      & --         \\
Si I & 614.2483  & -1.480  & ED            & 3.31    & 7.23       & 3.64    & 7.34      & 3.87    & 7.24       & 4.80    & 7.56       & 3.63    & 7.40       \\
Si I & 614.5016  & -1.430  & ED            & --      & --         & 4.44    & 7.43      & 3.30    & 7.09       & 4.83    & 7.51       & 4.21    & 7.46       \\
Si I & 615.5134  & -0.770  & ED            & 7.01    & 7.11       & 8.88    & 7.55      & 5.99    & 6.87       & 8.67    & 7.51       & 7.88    & 7.46       \\
\\
Ca I & 551.2980  & -0.290  & NBS            & 11.31   & 5.99       & 11.47   & 6.33      & 8.98    & 5.52       & 10.24   & 5.92       & 12.62   & 6.38       \\
Ca I & 585.7451  & 0.230   & NBS            & 15.84   & 6.08       & --      & --        & --      & --         & --      & --         & --      & --         \\
Ca I & 586.7562  & -1.610  & ED             & --      & --         & 4.19    & 6.30      & 4.00    & 6.04       & 5.12    & 6.32       & 4.98    & 6.33       \\
Ca I & 616.1297  & -1.020  & NBS            & 9.85    & 5.93       & 10.93   & 6.50      & 8.16    & 5.62       & 10.72   & 6.23       & 8.64    & 5.92       \\
Ca I & 616.6439  & -0.900  & NBS            & 9.91    & 5.82       & 10.05   & 6.21      & 7.21    & 5.36       & 10.42   & 6.05       & 9.61    & 5.99       \\
Ca I & 616.9042  & -0.550  & NBS            & 11.75   & 5.75       & 11.00   & 6.05      & --      & --         & --      & --         & 11.73   & 6.05       \\
Ca I & 645.5558  & -1.350  & NBS            & 8.84    & 6.09       & 7.95    & 6.24      & 8.75    & 6.030      & 9.46    & 6.30       & 9.08    & 6.34       \\
Ca I & 649.9650  & -0.590  & NBS            & 11.51   & 5.72       & 11.51   & 6.16      & --      & --         & 11.53   & 5.92       & 12.37   & 6.20       \\
Ca I & 650.8850  & -2.110  & NBS            & 3.29    & 6.04       & 3.21    & 6.13      & --      & --         & --      & --         & 2.90    & 5.93       \\
Ca I & 679.8479  & -2.320  & K88            & --      & --         & --      & --        & --      & --         & --      & --         & 1.79    & 6.06       \\
\\
\hline
\\
\multispan{10}{ED    Edvardsson, B. et al. 1993, A. and A. 275, 101 \hfill}\\
\multispan{10}{FF    Froese Fischer, C. 1975, Can.J.Phys. 53, 184; p. 338; 1979, JOSA 69, 118. \hfill}\\
\multispan{10}{FMW   Fuhr, J.R., Martin, G.A., and Wiese, W.L. 1988. J.Phys.Chem.Ref.Data 17, Suppl. 4. \hfill}\\
\multispan{10}{GARZ  Garz, T. 1973, a. and A. 26, 471. \hfill}\\
\multispan{10}{K88   Kurucz, R.L. 1988, Trans. IAU, XXB, M. McNally, ed., Dordrecht: Kluwer, 168-172. \hfill}\\
\multispan{10}{KZ    Kwiatkowski, M., Zimmermann, P., Biemont, E., and Grevesse, N. 1982, A. and A. 112, 337-340.}\\
\multispan{10}{LZ    Lincke, R. and Ziegenbein, G. 1971, Z. Phyzik, 241, 369. \hfill}\\
\multispan{10}{NBS   Wiese, W.L., Smith, M.W., and Glennon, B.M. 1966, NSRDS-NBS 4. \hfill}\\
\end{tabular}
}
\end{center}
\end{table*}

\begin{table*}
\caption{Coordinates and atmospheric parameters}
\label{coord}
\begin{center}
\begin{tabular}{lrrrrrrrrrrr}
\hline
\\
Star$^a$ & \multispan3{\hfill$\alpha(2000)^b$\hfill} &  \multispan3{\hfill$\delta(2000)^b$\hfill}&
\multispan2{$V~ (V-I)_0^c$} &
$\rm T_{eff}$ &log g & $\xi$  \\
\\
\hline
\\
432      &18& 53& 50.75&  $-30$& 27& 27.3&  17.55 & 0.965  & 4818 &2.30&1.3\\
628      &18& 53& 47.91&  $-30$& 26& 14.5&  18.00 & 0.928  & 4904 &2.50&2.0\\
635      &18& 53& 51.05&  $-30$& 26& 48.3&  18.01 & 0.954 & 4843 & 2.50&1.8\\
656      &18& 53& 45.71&  $-30$& 25& 57.3&  18.04 & 0.882 & 5017 & 2.50&1.6\\
709      &18& 53& 38.73&  $-30$& 29& 28.5&  18.09 & 0.917 & 4930 & 2.50&1.5\\
716      &18& 53& 52.97&  $-30$& 27& 12.8&  18.10 & 0.902 & 4967 & 2.50&2.0\\
717      &18& 53& 48.05&  $-30$& 29& 38.1&  18.10 & 0.872& 5042  & 2.50&1.3\\
772$^d$  &18& 53& 48.13&  $-30$& 32&  0.8&  18.15 & 0.947& 4891  & 2.50&1.5\\
867      &18& 53& 53.02&  $-30$& 27& 29.2&  18.30 & 0.933& 4892  & 2.50&2.0\\
879$^e$  &18& 53& 48.59&  $-30$& 30& 48.7&  18.33 & 0.965& 4891  & 2.50&1.4\\
894      &18& 53& 36.84&  $-30$& 29& 54.1&  18.34 & 0.940& 4876 & 2.50 &1.4\\
927      &18& 53& 51.69&  $-30$& 26& 50.7&  18.39 & 0.937& 4880 & 2.75 &1.2\\
\\
\hline
\end{tabular}
\\
\end{center}
\hbox{$^a$ see note to Table \ref{abund0}\hfill}
\hbox{$^b$  accurate to 0\farcs{3} (Ferraro \& Monaco 2002, private communication)\hfill}
\hbox{$^c$ The adopted reddening is $E(V-I)=0.22$ }
\hbox{$^d$ this is star  [BHM2000] 143  of \citet{B00}\hfill}
\hbox{$^e$ this is star  [BHM2000] 139  of \citet{B00}\hfill}
\end{table*}

\begin{table*}

\caption{Abundances}
\label{abund0}
\begin{center}
\begin{tabular}{lllclclclclc}
\hline
\\
Star$^a$ & S/N & A(FeI) & $n $& A(FeII) & $n$ & A(Mg) & $n$ & A(Si)& $n$ & A(Ca) & $n$ \\
         & @530nm &     &     &         &     &       &     &      &     &       &     \\
\hline
\\
Sun      & & 7.50 & & 7.50&  & 7.58&  & 7.55& &  6.36& \\
432      & 28 & $6.67\pm0.12$ & 12&    $6.75\pm0.13$ &8  & $ 6.74\pm 0.01$ & 2& $6.66\pm 0.16$ & 4 & $5.29\pm0.14$ & 6 \\ 
628      & 37 & $7.28\pm0.11$ & 9 & $7.43\pm0.14$ & 5 & $7.30\pm0.15$ & 4 & $7.31\pm0.11$ & 3 & $5.92\pm0.11$ & 6     \\
635      & 19 & $7.17\pm0.12$ & 10 & $7.13\pm0.19$ & 5 & $7.20\pm0.14$ & 4 & $7.08\pm0.12$ & 4 & $5.77\pm0.13$ & 11    \\
656      & 24 & $7.33\pm0.10$ & 14 & $7.35\pm0.13$ & 6 & $7.17\pm0.08$ & 3 & $7.29\pm0.07$ & 4 & $6.01\pm0.14$ & 9      \\
709      & 41 & $7.48\pm0.20$ & 15 & $7.41\pm0.11$ & 7 & $7.31\pm0.13$ & 4 & $7.34\pm0.08$ & 5 & $6.12\pm0.17$ & 11      \\
716      & 36 & $7.38\pm0.11$ & 12 & $7.33\pm0.11$ & 4 & $7.22\pm0.05$ & 3 & $7.26\pm0.14$ & 3 & $5.92\pm0.14$ & 8      \\
717      & 20 & $7.59\pm0.10$ & 12 & $7.49\pm0.16$ & 6 & $7.51\pm0.06$ & 3 & $7.49\pm0.11$ & 4 & $6.24\pm0.13$ & 8      \\
867      & 20 & $6.94\pm0.19$ & 12 & $7.13\pm0.11$ & 4 & $6.89\pm0.14$ & 4 & $7.00\pm0.17$ & 5 & $5.72\pm0.28$ & 5       \\
894      & 34 & $7.46\pm0.14$ & 13 & $7.35\pm0.13$ & 7 & $7.20\pm0.15$ & 3 & $7.51\pm0.04$ & 4 & $6.12\pm0.17$ & 6      \\ 
927      & 43 & $7.47\pm0.15$ & 15 & $7.35\pm0.06$ & 5 & $7.26\pm0.17$ & 3 & $7.42\pm0.07$ & 5 & $6.12\pm0.17$ & 9       \\
\\
\hline 
\\
\multispan{10}{$^a$ Star numbers are from \citet{Marconi} field 1, available
through CDS\hfill}\\
\multispan{10}{ at {\tt cdsarc.u-strasbg.fr/pub/cats/J/A+A/330/453/sagit1.dat}\hfill}\\
\end{tabular}
\end{center}
\end{table*}

\section{Analysis}

  Our analysis was performed on the spectra reduced 
with the UVES pipeline
delivered together with the raw data.
Our previous experience \citep{boni02} has shown that
the quality of this data is adequate for the
determination of abundances. 
The observed  radial velocity of each spectrum
was measured by cross-correlation with  a synthetic
spectrum.
This was done independently
for
both spectral ranges, corresponding to the two
CCDs of the mosaic, and was always consistent between the two CCDs.
The spectra were then  doppler shifted to
rest wavelength and different spectra of the same star
were  rebinned to the same wavelength step
and coadded.
The resulting S/N ratio of the coadded spectra is in the range
19 -- 43 at 530 nm and greater
at longer wavelengths. 
The equivalent widths listed in Table \ref{abund1} of individual lines
were measured using the {\tt iraf } task {\tt splot},
either by fitting a single gaussian or multiple
gaussians when the lines were slightly blended.
The standard
deviation of repeated measurement of the same line 
was of the order of the error estimated from
the Cayrel formula \citep{cayrel}, as expected.

The effective  temperatures listed in Table \ref{coord} were derived from the
$(V-I)_0$ colour through the calibration of \citet{alonso},
we adopted the reddening $E(V-I)=0.22$, $A_V = 0.55$
and a distance
modulus $m-M=16.95$ from \citet{Marconi}.
For each star we computed a model atmosphere
using version 9 of the ATLAS code \citep{k93}
with the above \teff ~and log g = 2.5, which was estimated
from the location of the stars in the $(V-I)_0, M_v$ diagram
and the isochrones of \citet{straniero} of ages between  8 and 10Gyr,
which is an age range compatible with
the anaysis of \citet{Marconi}.
The abundances were determined
using the WIDTH code \citep{k93} except for oxygen; since the
only oxygen indicator available is the [OI] 630nm line, which is 
blended with the Ni I 630.034 nm line
we used spectrum synthesis with the SYNTHE
code  \citep{k93}. 
For this blending line we used the recently measured log gf = --2.11 
of \citet{j03}, and assumed
that Ni scales with Fe.
We re-determined the oxygen abundance also for star \# 772 and
the upper limit for \# 879, 
already studied by \citet{B00}, with this new log gf for the Ni I 
line.  
The model atmospheres were computed switching off the 
overshooting option, according to the prescription of
\citet{castelli} and an opacity distribution function (ODF)
with microturbulent velocity $\xi=1$ \kms,
suitable metallicity and [$\alpha$/Fe]= 0.0 or +0.4,
adjusted iteratitively in the course of the analysis.

For two stars we decided, in the course of the analysis,
to change the surface gravity in order to have a better
Fe I /Fe II ionization equilibrium. The new gravity is consistent
with the position of the stars in the colour magnitude diagram.
Several stars turn out to have solar metallicity and
[$\alpha$/Fe]$<$0. 
Since the effect of different [$\alpha$/Fe] ratios
on the temperature structure of model-atmospheres
is higher at higher metallicities,  our
computations, performed with models computed with ODFs
with [$\alpha$/Fe]=0., are inconsistent.
Fiorella Castelli kindly  computed at our request an ODF
with solar metallicity and [$\alpha$/Fe] =--0.2 
using the same input data as in \citet{castelli02}
and we computed models using this new ODF.  
The effect of the new models on our derived abundances
was tiny, of the order of a few hundredths of dex.
This is due to the fact that the lines we employ are
either of medium strength or strong and are therefore formed
in rather superficial layers. The difference
in structure due to different [$\alpha$/Fe] ratios
is strongest at large depths, so that only abundances
derived from very weak lines should be affected.

The results of our analysis are summarized in Table
\ref{abund0}. In Table  \ref{abund} we provide
the [X/Fe] ratios for all of the Sgr giants observed with UVES,
thus including also the two  stars from \citet{B00}.

The O abundance depends also on the, presently unknown, C abundance,
since a significant fraction of O is locked as CO. We assumed 
[C/Fe]=0.0 . For carbon abundances in the range
$-0.5 \le $ [C/Fe]$\le +0.5$  the effect on the derived
O abundance is of the order of 0.01 dex and may therefore be
neglected. We may exclude significant ($> 0.5 $dex) C enhancement,
since our spectrum synthesis suggests we would detect some CN lines;
this is in agreement with \citet{BC03}, whose automatic code did not flag
any of the stars as ``suspect CN''.
We cannot exclude the C-starved case ([C/Fe]$\la -2.0$), 
which could be observed
in giant stars if the surface material has been polluted by CN cycled
material in which C has been depleted. In Galactic globular
clusters carbon abundances are in fact usually found below
their oxygen abundances \citep{brown90}. However the C-starved
case would imply {\em lower} [O/Fe] ratios
of at most 0.2 dex. Since the hypothesis
[C/Fe]=0.0 provides [O/Fe] ratios 
which are in line with the ratios of other $\alpha$ elements to iron,
we think it is a reasonable working hypothesis. 
Furthermore it is extremely unlikely that all stars in the sample
are C-starved stars.
When C abundances will
be derived for these stars the O abundances might be re-considered.
     
In order to provide a reference for differential
metallicities, if desired, we determined the Fe abundance for
two giants in the Hyades: HD 27371 and HD 27697. These stars
are almost twins and have atmospheric parameters close to those
of the Sgr stars analysed in the present paper. A differential
metallicity with respect to these should therefore be independent
of our adopted temperature scale, model atmospheres and atomic data.
We measured equivalent widths on the FEROS spectra for 12 Fe I and
1 Fe II lines out of those in Table \ref{abund1}.
There could be some concern in comparing equivalent widths measured
with two different spectrographs, of slightly different  resolution.
However scattered light is known to be very low in both spectrographs
and we believe this to be a minor problem
\citep{kaufer}.

We used the measurements of  \citet{Johnson}, as 
reported in the General Catalog of
Photometric data \citep{GCPD},
which imply  $V-I=1.20$ for both stars, and a zero 
reddening, to derive an effective temperature of 4880 K for both stars.
We assumed log g = 2.50, like for our program stars
and computed a model atmosphere with these parameters,
solar metallicity and microturbuence of 1 kms$^{-1}$. 
The derived [Fe/H] is reported in Table \ref{hyades}
together with other determinations for the two stars
found in the catalog of \citet{giusa}.
Our determination is compatible, within errors with all
previous determinations. It is interesting to notice that 
with our parameters the two stars have the same [Fe/H],
as is expected from the similarity of the two spectra.
The FeI/FeII  ionization balance is achieved
within errors, for our adopted gravity, albeit Fe II
is represented by a single line.

The line to line scatter provided in Table \ref{abund0}
provides a good estimate of the statistical error arising 
from the noise in our spectra and uncertainties in the measurement
of equivalent widths; the errors in the mean abundance may be estimated
assuming that each line provides an independent measure
of the abundance and dividing the  the dispersion by $\sqrt{n}$, where 
$n$ is the number of measured lines.
To this one should add linearly the errors arising from the uncertainties in 
the atmospheric parameters.
As a representative star we take star \# 628
and report in Table \ref{errors} these errors.

\begin{table*}

\caption{Errors in the abundances of star \# 628 due to uncertainties
in the atmospheric parameters}
\label{errors}
\begin{center}
\begin{tabular}{lrrrrrrrr}
\hline
\\
 & $\Delta$A(FeI) & $\Delta$A(FeII) &$\Delta$A(O) & $\Delta$A(MgI) & $\Delta$A(SiI) & $\Delta$A(CaI) \\
\hline
\\
$\Delta\xi =  \pm 0.2$ \kms &  $^{-0.06}_{+ 0.07}$ & $^{-0.04}_{+ 0.05}$ & $\mp 0.05$ & $\mp 0.04$ & $^{-0.03}_{+ 0.04}$ &  $^{-0.05}_{+ 0.06}$\\
\\
$\rm \Delta T_{eff} = \pm 100$ K &$ ^{+0.14}_{-0.06}$ & $^{-0.06}_{+0.12}$ & $\mp 0.01$ & $^{+0.09}_{-0.06}$ & 
$^{+0.00}_{-0.05} $ & $^{+0.17}_{-0.10} $ \\  
\\
$\Delta \log g = \pm 0.50 $ & $^{+0.05}_{-0.03}$ & $^{+0.29}_{-0.20}$ & $^{+0.25}_{-0.24}$ & $^{-0.03}_{+0.01}$
& $^{+0.11}_{-0.06}$ & $^{-0.03}_{+0.01}$
\\
\\
\hline
\\
\end{tabular}
\end{center}
\end{table*}

\begin{table*}
\caption{Abundance ratios}
\label{abund}
\begin{center}
\begin{tabular}{lrrrrr}
\hline
\\
Star$^a$    & \multispan1{\hfill[Fe/H]\hfill}& [O/FeII]  &  [Mg/Fe] & [Si/Fe] & [Ca/Fe]\\ 
\\
\hline
\\
432 &    $-0.83\pm 0.12$ &$+0.01$  & $-0.01$ & $-0.06$ &  $-0.24$   \\
628 &    $-0.22\pm 0.11$ &$-0.16$  & $-0.06$ & $-0.02$ &  $-0.22$   \\
635 &    $-0.33\pm 0.12$ &$-0.02$  & $-0.05$ & $-0.14$ &  $-0.26$   \\
656 &    $-0.17\pm 0.10$ &$-0.18$  & $-0.24$ & $-0.09$ &  $-0.18$   \\
709 &    $-0.02\pm 0.20$ &$-0.14$  & $-0.25$ & $-0.21$ &  $-0.22$   \\
716 &    $-0.12\pm 0.11$ &$-0.16$  & $-0.24$ & $-0.17$ &  $-0.32$   \\
717 &    $+0.09\pm 0.11$ &$-0.09$  & $-0.16$ & $-0.15$ &  $-0.21$   \\
772$^b$& $-0.21\pm 0.19$ &$-0.11$  & $-0.23$ & $-0.07$ &  $-0.26$\\   
867    & $-0.56\pm 0.19$ &$-0.01$  & $-0.13$ & $+0.01$ &  $-0.08$\\   
879$^c$& $-0.28\pm 0.16$ &$\le 0.18$& $-0.05$ & $-0.07$ &  $-0.21$\\  
894 &    $-0.04\pm 0.14$ &$+0.11$  & $-0.34$ & $+0.00$ &  $-0.20$   \\
927 &    $-0.03\pm 0.15$ &$-0.09$  & $-0.29$ & $-0.10$ &  $-0.20$   \\
\\
\hline
\end{tabular}
\\
$^a$ see note to Table \ref{abund0}\\
$^b$ this is star  [BHM2000] 143  of \citet{B00}\\
$^c$ this is star  [BHM2000] 139  of \citet{B00}\\
\end{center}
\end{table*}

\begin{figure}
\centering
\resizebox{\hsize}{!}{\includegraphics[clip=true]{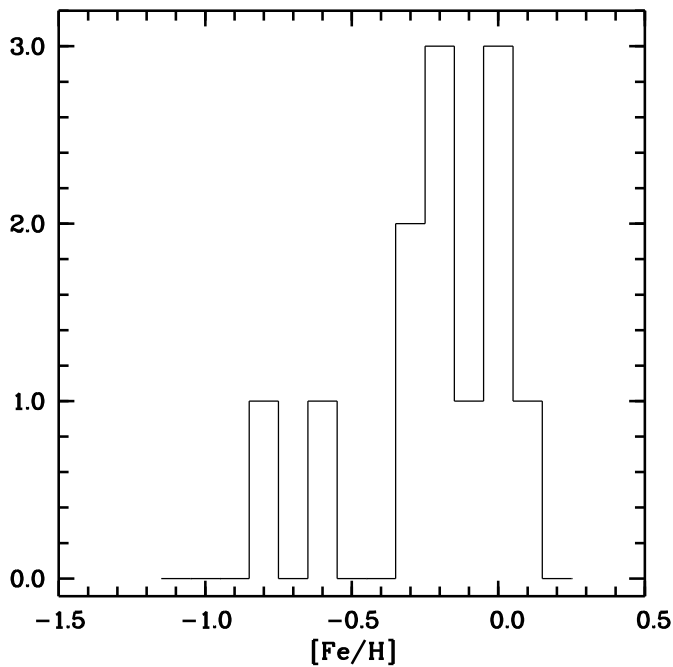}}
\caption{The metallicity distribution of the 10 Sgr giants of
the present paper and the two of \citet{B00} }
\label{histofe}
\end{figure}

\begin{table*}

\caption{Iron abundances of two Hyades giants}
\label{hyades}
\begin{center}
\begin{tabular}{rlllllllllll}
\hline
\\
\multispan6{\hfill HD 27371\hfill}&\multispan6{\hfill HD 27697\hfill}\\
\\
\hline
\\
Ref. & \teff  & log g & $\xi$ & \hfill [FeI/H] \hfill & [FeII/H] & 
Ref. & \teff  & log g & $\xi$ & \hfill [FeI/H] \hfill & [FeII/H] \\
\hline
\\

1 & 4880 & 2.50 & 1.3 & $+0.21\pm 0.13$ & +0.15 & 1 & 4880 & 2.50 & 1.4 &
$+0.23\pm 0.13$ & +0.05\\ 
2 & 5143  & 3.02 &    &  +0.07          &       & 2  & 5143 & 2.98 &  & +0.08\\
3 & 5040  & 2.58 &    &  +0.17          &       & 3  & 5040 & 2.44 &  & +0.41\\
4 & 5663  & 2.0  &    &  $-0.26$          &       & 4  &                       \\
5 & 4271  & 3.0  &    &  +0.34          &       & 5  & 4235  & 3.0  & &  +0.30\\
6 & 4990  & 2.80 &    &  +0.20          &       & 6       \\
7 & 4800  & 2.60 &    &  +0.09          &       & 7  \\
8 & 4930  & 2.90 &    &  $-0.02$          &       & 8  & 4940  & 2.85 & &  +0.00\\
9 & 4900  & 2.60 &    &  +0.13          &       & 9  & 4875  & 2.40 & &  +0.06\\
10& 4965  & 2.65 &    &  $+0.15\pm 0.03$ &       \\
\hline
\\
\end{tabular}
\noindent \parbox[t]{15cm}{\small
(1) This paper, (2) \citet{lambert}, (3) \citet{gratto82},
(4) \citet{Koma80}, (5) \citet{Koma85}, (6) \citet{gratto86},
(7) \citet{Fernan}, (8) \cite{mc90}, (9) \citet{Luck},
(10) \citet{Smith}}
\end{center}
\end{table*}

\section{Metallicity distribution}

The histogram  of the [Fe/H] for the 12 stars observed with UVES
is shown in Fig. \ref{histofe}. 
In spite of the small number of stars two features are obvious:
1) the bulk of the population is metal-rich with $\rm -0.5\la [Fe/H]\la 0.0$;
2) the spread is of the order of 1 dex.

It is clear that it is not easy to evaluate the
relative contribution of  each metallicity bin to the 
total population of Sgr
from a sample of only 12 stars. Here and in the rest of the paper
we call ``dominant population'' the one which seems
to dominate   in Fig. \ref{histofe}, i.e. the stars with
[Fe/H]$> -0.5$. When accurate metallicities for a
larger number of stars become available it will be interesting
to see if the picture which emerges from this
small sample will persist.

\begin{figure}
\centering
\resizebox{\hsize}{!}{\includegraphics[clip=true]{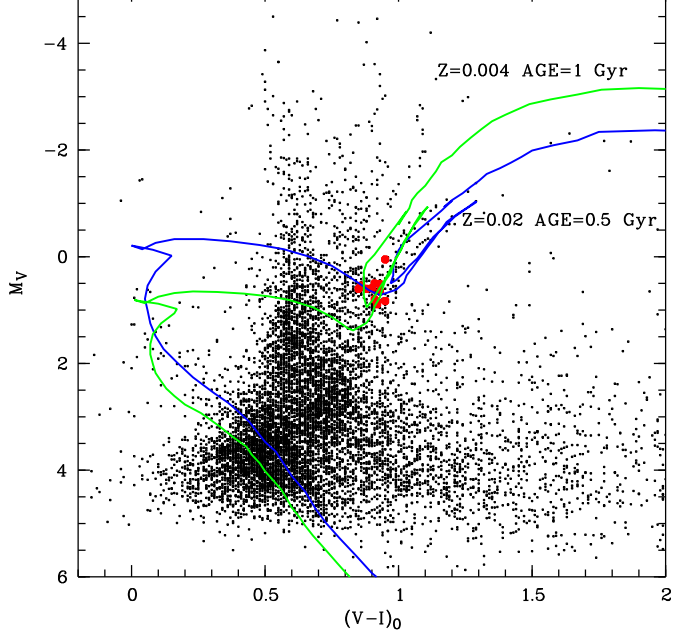}}
\caption{The colour--magnitude diagram of {\sl field 1}of \citet{Marconi},
the reddening adopted is $E(V-I)=0.22$ , $A_V=0.55$ and
the distance modulus $(m-M)=16.95$, from the same reference.
The stars observed by UVES are shown with larger symbols, two Padova
isochrones \citep{girardi}, 
covering the range in $Z$ found for the program stars are shown. }
\label{cmd}
\end{figure}

Our results are somewhat at odds with some of the photometric
estimates of these quantities.
In Fig. \ref{cmd} we show the colour-magnitude diagram (CMD)
for the field, where our stars were selected, from
\citet{Marconi}. The 12 stars observed with UVES are highlighted
as bigger filled symbols.
These stars sample the RGB at roughly constant luminosity,
spanning all of its  width.
From this plot it seems likely that, 
in spite of its limited size,
our sample captures the whole metallicity spread in
this field of Sgr. 
We may estimate the corresponding spread in $Z$:
our measures indicate
that  O traces the other
$\alpha$ elements and we assume C and N trace Fe,  
then for our program stars we find $0.002\la Z\la 0.014$.
Note, that because of the  lack of enhancement of $\alpha$ elements,
even the stars with $\rm [Fe/H]\sim 0.0$ have a $Z$
which is slightly sub-solar.
In Fig. \ref{cmd} also two Padova isochrones \citep{girardi}
for the $Z$ values relevant to the program stars are shown.
Two things are worth noting:
\begin{enumerate}
\item      to reproduce the observed RGB of Sgr
        with isochrones of this high $Z$
 very young ages,
      of the order of 0.5 to 1.0 Gyr, are required
\item the "blue plume", which is obvious in the CMD,
      occupies the same region in the CMD
     as   the Main Sequence
       and Turn-Off of this metal-rich population.
\end{enumerate}

If the "blue plume" did not exist, the extremely
young age we suggest for  the dominant population
of this region of Sgr, would not be supported.
Instead the fact that this  explains a feature
of the CMD, which we were not seeking to explain,
reinforces our interpretation.
A possible problem is that there seem to be too few
blue stars to explain all of the RGB stars at $M_V \sim 0.9$.
A full model population study of the CMD of Sgr is beyond
the scope of this paper, as is a precise
determination of the age(s) of the metal-rich
population of Sgr.
In any case, in the interpretation 
of the CMD, one will have  to take into account also:
\begin{enumerate}
\item possible incompleteness of the photometry at the
      luminosity of the ``blue plume'';
\item contamination at the apparent luminosity of the RGB
      by the Bulge and the Galactic disc;
\item the fact that if  several
      populations of different ages are present
      each will contribute some stars at  $M_V \sim 0.9$.
\end{enumerate}

The young age of the Sgr dominant population
also explains the disagreement in metallicity
spread and mean metallicity between the
spectroscopic analysis and the photometric
estimate.
\citet{Marconi} compared the CMD shown
in Fig. \ref{cmd} with the fiducial
lines of two Galactic Globular Clusters:
47 Tuc and M2.
The age of M2 is 2 Gyr older than that
of M3 \citep{lee} which is 11.3 Gyr old 
\citep{salaris} therefore $\sim 13.3$ Gyr.
The age of 47 Tuc is 12.5 Gyr, according to 
\citet{carretta}.
Both clusters are representative of an old population.
We are thus in presence of a rather extreme
case of the well known
age--metallicity degeneracy.
With our assumed ages and spectroscopic
metallicities the morphology of the CMD  
is well explained by the theoretical
isochrones.
The high metallicity 
derived by \citet{B00} raised some concern
among several workers in the field. 
For example \citet{monaco}
claimed that the high metallicity
derived by \citet{B00} is not representative
of the bulk of Sgr and took
the metallicity distribution
of \citet{mcwilliam}
as being in better agreement with the photometric
data. Our results instead
reinforce the original findings of \citet{B00}.
Morever the metallicity distribution
found by \citet{mcwilliam} is quite
similar to ours except for the 3 truly
metal-poor stars, out of their sample of 14; 
analogs to these are totally lacking in our sample. 
It is relevant to point out that 
\citet{monaco}, exactly like \citet{Marconi}, base
their conclusions 
on comparison with fiducial lines of Galactic
Globular Clusters(GGCs). On the other hand \citet{cole},
who made use  of the slope of the RGB in an IR colour
magnitude diagram, arrived at an 
estimate of [Fe/H]=$-0.5\pm 0.2$ for the bulk
of the Sgr population. Quite correctly \citeauthor{cole}, 
pointed out that this estimate could in fact be too low,
since the relations are
calibrated on GGCs and age effects may bias 
downwards the results by 0.1-0.2 dex; furthermore the lack
of $\alpha-$ enhancement in Sgr (confirmed by the present
analysis), at variance with GCCs, would also act in the direction
to lower the photometric metallicity estimate.

Therefore we believe that the high mean metallicity
we derive is entirely consistent with extant photometric
data, once a young age is allowed for.
The real issue, concerning our metallicity distribution,
is the lack of the very metal poor population found by
\citet{mcwilliam}. For this we see only two
possible explanations: either the metal-poor population
is not present in the region of Sgr studied by us
or our target selection criteria have been such
that we missed it.

In the first place  it must be borne in mind, that 
we and \citeauthor{mcwilliam}
are sampling both different regions
of Sgr {\em and}
different evolutionary phases. Our stars are {\it bona fide}
first ascent red giants or clump stars, while the sample of
\citeauthor{mcwilliam} quite likely contains also some AGB stars.
Unfortunately \citeauthor{mcwilliam} do not provide
coordinates for their stars, yet they say that they
were selected from the fields imaged
by \citet{SL}.
These two fields are centered on M54 and 12 arcmin North
of the cluster center.
On the other hand our stars lie in {\sl field 1}
of \citet{Marconi}, which is $\sim 22$ arcmin West
of the center of M54.
We recall that \citet{Marconi} found indeed a difference
in colour at the base of the RGB between M54 and {\it field1}
and suggested as most likely explanation a difference
in mean metallicity of the order of 0.5 dex.
We suspect that the metal--poor stars found
by \citeauthor{mcwilliam} either belong to M54
or to the neighbouring field (possibly M54 debris) which
shares the same metallicity. If this were the case
the similarity between the metal--poor
stars of \citeauthor{mcwilliam} and the M54
giants investigated by \citet{brown} is not surprising.
A possible solution is therefore that the metal-poor population
is missing in the region sampled by us.

Inspection of Fig. \ref{cmd} shows that to the blue
of the 1 Gyr isochrone there is an RGB which, if attributed to
Sgr, could indeed represent a metal-poor population. In our
low-resolution survey, aimed at selecting radial velocity
members of Sgr, we ignored that RGB, attributing it to the Bulge.
Radial velocity measurements could ascertain if this is the case
or if it belongs to Sgr instead.  If this were the case then
the lack of detection of the metal-poor population can be attributed
to our selection of stars to be observed spectroscopically.

The existence of a metal-poor population in Sgr is not  questionable,
given the results of 
\citet{brown} for M54, of \citet{mcwilliam} for the Sgr field.
\citet{cs00} and \citet{cs01} have found  RR Lyr variables
in Sgr with a period distribution which points towards a mean metallicity, 
for this old metal-poor population, of [Fe/H]$\sim -1.6$. 
The real issue is the relative contribution of the various 
populations and their spatial extents.

\begin{figure}
\centering
\resizebox{\hsize}{!}{\includegraphics[clip=true]{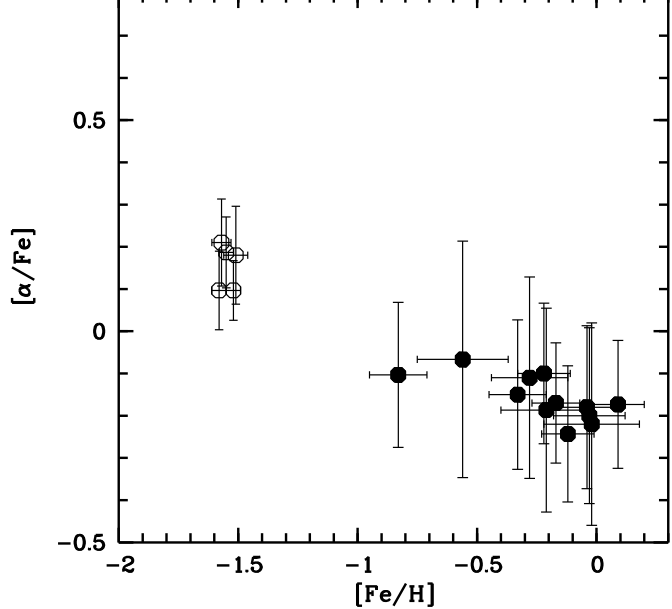}}
\caption{The [Fe/H], [$\alpha$/Fe] diagram for the programme stars (filled
symbols. The 5 stars of M54 studied b \citet{brown} are shown as open
symbols. [$\alpha$/Fe] in this plot
is defined as $1\over 3$([Mg/Fe]+[Si/Fe]+[Ca/Fe]).}
\label{plalpha}
\end{figure}

\begin{figure}
\centering
\resizebox{\hsize}{!}{\includegraphics[clip=true]{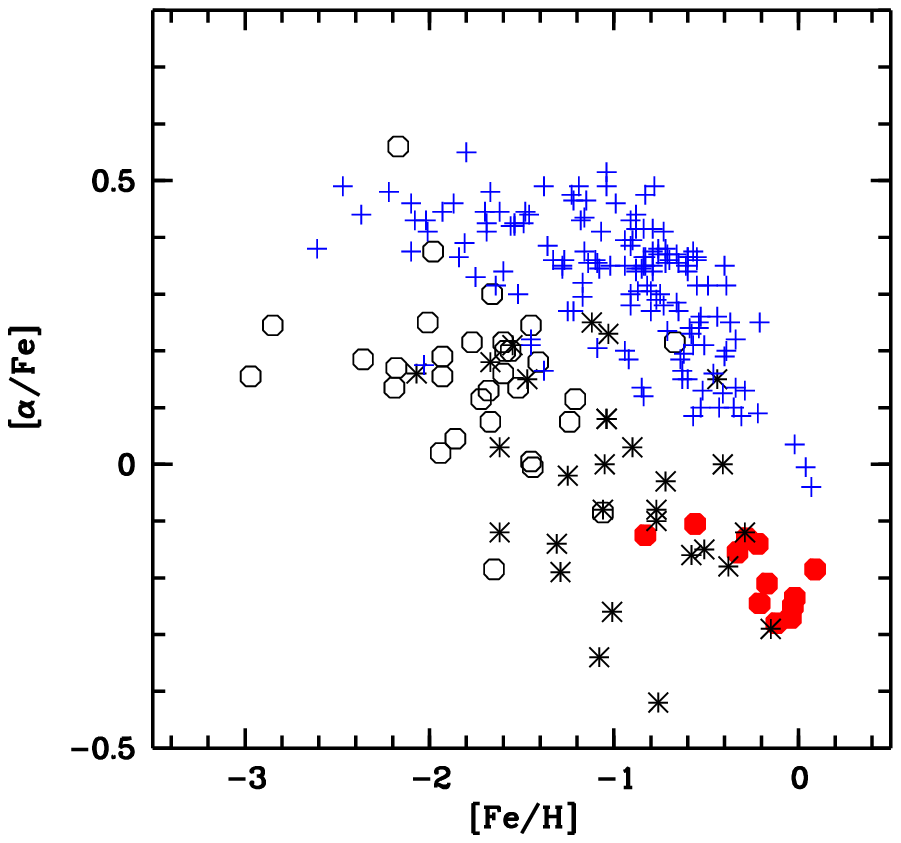}}
\caption{The [Fe/H], [$\alpha$/Fe] diagram for the programme stars (filled
symbols), the stars
in the Local Group  dwarf spheoridals, 
Draco, Ursa Minor, Sextans from \citet{shetrone}, Carina, Sculptor,
Fornax and Leo  from \citet{Shetrone03} (open circles),
Galactic stars from \citet{gratton03} (crosses) and for DLAs from
\citet{centurion} (asterisks). For the stars [$\alpha$/Fe], in this plot,
is defined as $0.5\times$([Mg/Fe]+[Ca/Fe]), for the DLAs
[Si/Zn] is used as a proxy for [$\alpha$/Fe] and [Zn/H]
as a proxy for [Fe/H]
}
\label{plalpha_dla}
\end{figure}

\section{$\alpha$/Fe ratios}

The ratios of $\alpha$ elements to iron peak elements
allow  one to place important constraints on the star formation
history of a galaxy. The common wisdom is that $\alpha$
elements are  synthesized by core collapse SNe, while
Fe is synthesized mainly  by Type Ia SNe  \citep[e.g.][~and
references therein]{marconi94}.
At early time the gas is enriched by SNeII only, thus the $\alpha$/Fe
ratio is that characteristic of the yields of massive stars.
As time proceeds the type Ia SNe begin to contribute to the Fe
content, lowering the $\alpha$/Fe ratio.
In the Milky Way at low metallicities the  $\alpha$/Fe ratio is
higher than at solar metallicity. As shown in figure 3 of
\citet{marconi94} a slow, {\em bursting} or {\em gasping} star formation rate
may attain a solar $\alpha$/Fe ratio at considerably lower metallicities,
and even reach values {\em lower} than the solar ratio.

In Fig. \ref{plalpha} we show the mean [Fe/H], [$\alpha$/Fe] diagram
for the programme stars. As  [$\alpha$/Fe] we have adopted the straight
mean of [Mg/Fe], [Si/Fe] and [Ca/Fe]; oxygen has not been included in the mean
since the errors associated are considerably larger than for 
the other elements, note however that in all cases [O/Fe]
is consistent with the above defined  [$\alpha$/Fe].
The error on this this ratio was computed by propagating the errors
on the various [X/H] ratios involved.
The figure provides an   indication
that the  $\alpha$/Fe ratio is below solar, although the errors are such
that [$\alpha$/Fe]=0.0 cannot be excluded. What can confidently be 
excluded is an $\alpha$ enhancement, even at the lowest metallicities
observed, which strongly suggests a slow or bursting 
star formation rate. 
Low $\alpha$/Fe ratios
seem to characterize also the other dwarf spheroidals for which 
spectroscopic abundance ratios are available \citep{shetrone,Shetrone03}.
Also the metal rich part of the sample of \citet{mcwilliam} shows
low  $\alpha$/Fe ratios, consistent with ours, 
within errors.
The data of \citeauthor{mcwilliam} gives the impression of an increase
of the $\alpha$/Fe ratio with decreasing metallicity, however this
impression is driven only by data for the three most metal-poor stars.
Our data instead seems very  flat, with a very weak trend in $\alpha$/Fe,
if any. 
The 5 stars of M54 studied by \citet{brown}, which have [Fe/H]$\sim -1.5$,
comparable to the most metal-poor stars found by \citet{mcwilliam}, show a
very mild $\alpha$ enhancement, of the order of [$\alpha$/Fe]=+0.2 or lower.
When joined with our data (Fig. \ref{plalpha}) they suggest a mild
increase in $\alpha$/Fe ratio, which is not so evident from our data
alone.  The $\alpha$ enhancement in M54 seems slightly lower than that
measured in the most metal-poor stars of \citet{mcwilliam}, although they
are compatible within errors.

In Fig. \ref{plalpha_dla} we assemble together our data
for Sgr, the data of \citet{shetrone} for the
Local Group dwarf spheoridals Draco, Ursa Minor, Sextans, 
and of  \citet{Shetrone03}  for  Carina, Sculptor,
Fornax and Leo I , the data of \citet{gratton03} on Galactic stars and the
data compiled by \citet{centurion} for Damped Lyman $\alpha$ systems (DLAs).
For all the stars in this plot [$\alpha$/Fe] has been defined as
the mean of [Mg/Fe] and [Ca/Fe]. This  is different
from the definition used in Fig. \ref{plalpha}
because most of the stars  of \citet{shetrone} and  \citet{Shetrone03} 
do not have any Si measurement. Comparison of our points in
figures \ref{plalpha} and \ref{plalpha_dla} shows that there
is little difference in the two definitions of [$\alpha$/Fe].
For the DLAs we have used [Si/Zn] as a proxy for [$\alpha$/Fe]
and [Zn/H] as a proxy for [Fe/H],  because Zn is unaffected
by dust depletion and Si is only mildly depleted in DLAs 
\citep{vladilo,centurion}.
In the metallicity range relevant to DLAs,
Zn tracks Fe in Galactic stars, possibly with a small offset of 
the order of 0.1 dex\citep{gratton03}; we stress that the picture
does not change significantly even taking this small offset
into account.
Figure \ref{plalpha_dla} clearly shows that dSph galaxies and most
of DLAs are on a chemical evolutionary path different from that
of the Milky Way. There seems to be continuity among the dwarf 
spheroidals which may suggest that they are all on similar evolutionary
paths, but some are more chemically evolved than others.
It is also tempting to conclude that most of the DLAs follow a 
chemical evolution similar to that of the dwarf spheroidals.
From there it takes only one more step to conclude that most DLAs {\em are}
dwarf spheroidals caught when they were still gas-rich.
The argument that most of DLAs are characterized by low star
formation rate and are therefore dwarf galaxies, has already been 
put forward by \citet{centurion00}. It receives support from
the few imaging studies, which show many of the DLAs to be
associated with dwarf galaxies and low surface brightness
galaxies \citep{lebrun,rao}. However up to now there was no clear
observational evidence that the $\alpha$ to iron ratios of dwarf
galaxies are indeed similar to those of DLAs, as expected from
chemical evolution models \citep{marconi94}, Fig. \ref{plalpha_dla} 
forcefully provides this evidence.

\section{A scenario for star formation history in Sgr}

We believe that our findings, when joined to the
other information we have of Sgr, suggest a well
defined star formation scenario, whose viability,
however, needs to be verified by hydrodynamical
simulations.

Sgr has undergone many star bursts, and each starburst
is triggered by the interaction of Sgr with the disc of our
Galaxy when Sgr passes through it.
The simulations of \citet{ir98} indicate that the HI disc
of the Galaxy is shocked and warped by the passage
of the dwarf galaxy. Furthermore there is substantial
heating in the disc shocks and in the tidal tails which 
follow the dwarf galaxy.
The simulations \citeauthor{ir98}  are not designed
to investigate star formation, however 
\citet{Ng2} points out that the number density and temperature
of the gas are such that a part of the gaseous material
should be converted to stars. 
It therefore seems plausible
that star formation is triggered by  the shock
both in the Galactic disc and in the dwarf galaxy, if it
possesses any gas. Besides it is conceivable that 
some of the
gas may be stripped from the Galaxy and follow Sgr,
perhaps forming stars.
Dedicated simulations, with a live
Halo, rather than the static one used by \citeauthor{ir98},
are highly desirable to verify if this scenario is actually
viable.

\subsection{Carbon stars}

According to this scenario
the dominant population we see in {\sl field 1} was
formed  when Sgr last passed through the 
Galactic disc. This scenario has also the virtue
of explaining the mystery of the Bulge C stars
\citep{Ng1,Ng2}.
The presence of carbon stars in fields of the Bulge
has been known for over twenty years \citep{blanco78,azzo88}. 
However if these stars belong
to the Bulge they are about 2.5 mag too faint
to be on the AGB tip, where the carbon-star
phenomenon is believed to take place. Moreover the Bulge population
is believed to be old, while for a solar metallicity
population the mass of the carbon stars
should be greater than 1.2 M\sun ~and their 
age younger than 4 Gyr \citep{marigo99}.
This fact lead \citeauthor{Ng1} to look for an alternate explanation.
He noticed that the distance modulus of Sgr is in fact 2.5 mag fainter
than that of the Bulge. This implies that, 
if placed at the distance of the Sgr, the Bulge carbon stars
appear to have the theoretically expected luminosity. 
However the radial velocities of the 
carbon stars of \citet{azzo91}  does not allow
to consider them as members of Sagittarius.
On the other hand Sagittarius surely has 
some  carbon which are established
radial velocity members
stars \citep{ibata95} and several candidate stars
\citep{wh96,ngs97}. 
All these observations are explained quite naturally by the
star formation scenario proposed by us: the last  crossing
of the Galactic disc by
Sgr triggered star formation both in Sgr (the metal-rich 
population and the blue plume) as well as in the Galactic
disc; some   of these stars evolved to become carbon stars
in the disc (which has been disturbed) at a distance
from us which is about that of Sgr and these are
the stars of \citet{azzo91} and all the other
so-called Bulge carbon stars; the same happened to some
stars in  Sgr and these are the carbon stars with
Sgr membership \citep{ibata95} and also the other candidate stars,
if confirmed \citep{wh96,ngs97}.
If this is case the difference in J-K colour between the two groups 
cannot be due to a difference in age, but only to
differences in metallicity and reddening. 

We note that overdensities of carbon stars in the Halo are one
of the main tracers of stellar ``streams'' ascribed to the
disruption of Sgr \citep{ibata01}.

\subsection{Was there gas in Sgr to support star formation ?}

A major problem with this scenario is the lack
of any detectable amount of gas in Sgr
\citep{Burton}.
One is forced to admit that during the last passage
the star formation and the gas stripping has exhausted
all the available gas. If that is so, no star formation
shall take place in the next passage, unless
Sgr is capable to capture gas from the Galactic disc
or to capture intergalactic gas clouds.
Such an event has been suggested to be the cause of the present
star forming activity in the Magellanic Clouds 
\citep{hira99,hira97}. 

We also wish to point out that a factor
that may help Sgr in retaining hot gas, thus providing
material suitable for forming the next generation of
stars, is the fact that 
the cooling time is of the order of, or even smaller
than, the crossing time. This is due to the high
metallicity which makes  the line cooling very efficient,
as may be deduced by 
following the treatment of 
\citet{hira99} and setting  
$\zeta = Z/Z_\odot=0.7$. 

\subsection{Similarity with the LMC}

There are some important similarities between Sgr and the LMC, which
seem to suggest that 
Sgr was more massive in the past and perhaps
its mass was of the order of that of the LMC: 
1) the RR Lyr populations  
of the two galaxies are basically indistinguishable; 2) so
is the mean metallicity of the younger populations ([Fe/H]$\sim -0.25$ );
3) at the intermediate and high metallicities also the LMC
shows low $\alpha$/iron ratios (Hill et al. in preparation);
4) the pattern of the abundances of neutron capture elements is similar,
as pointed out by \citet{B00}.

\section{Conclusions}

With the 10 stars analysed in this paper, the
sample of Sgr giants in {\em field 1 } of \citet{Marconi}
has risen to 12. The dominating population appears to be
close to solar metallicity and the metallicity range is
$\rm -0.8 \la [Fe/H] \la 0.0 $. The ratio of $\alpha$ elements
to iron is sub-solar or solar even at the lowest metallicity
observed. In our sample stars which are as
metal-poor as M 54 or  the most metal-poor stars in the
sample of \citet{mcwilliam} are missing. 
Therefore either such very metal-poor component  has a very low space density
in {\em field 1 }, perhaps is even absent, or our
selection criterion has totally missed this population.
This issue needs to be elucidated by the observation
of a larger number of stars in the field which
we plan to perform with the FLAMES facility on the VLT
\citep{pasquini,pasquini02}.

Our results raise an important  question: why is the
most metal-poor population of Sgr  observed by us
so (relatively)
metal rich ?
In our opinion there are two most likely mechanisms
for this enrichment:
1) the gas has been enriched by previous generations of
Sgr stars 
or
2) the gas out of which  stars of
Sgr were formed was polluted by SNe in the
Milky Way, the pollution process being possibly 
favoured by the passage of Sgr through the Galactic disc. 

Either solution has some problems. In the first
case  it is not clear where the low-mass
stars of the previous generations of Sgr are, nor is it
clear how Sgr managed to retain the SNe ejecta in order
to attain such a high metallicity. 
In the second case it is not clear how the pollution may
take place;  the passage of Sgr through the
Galactic disc might offer the opportunity, however 
the degree of pollution, if any, has still to be  
evaluated.

Whichever the case we note that the high metallicity
we derived places Sgr clearly outside the
metallicity - luminosity correlation which seems to hold
for other Local Group galaxies \citep[~and references therein]{vdb99}:
Sagittarius is underluminous for its metallicity.
The capability of attaining a high metallicity is usually
associated with the ability to retain the SNe ejecta and therefore
a rather large gravitational potential.
We therefore
believe that an explanation of the high metallicity
associated to a relatively low luminosity
must be sought among the following:
1) Sgr posseses an extraordinarily large amount
of dark matter ; 2) Sgr was much more massive in the past, during the
phase in which it raised its metallicity and has now
lost much of its mass due to interaction with the Milky Way; 
3) the interaction with the Milky
Way, through pollution and/or tidal interaction which resulted in
increased star formation activity and ability to retain the SNe ejecta. 
These are not
mutually exclusive and a combination of the above is possible.

\begin{acknowledgements}

The authors are grateful
to F. Castelli for computing a customized
ODF and for many interesting discussions
on the physics of stellar atmospheres.
We are also indebted to F. Ferraro and L. Monaco
for providing accurate coordinates for our program stars. 
PB also aknowledges helpful discussions with
L. Girardi and S. Zaggia.
This research was done with support from the
Italian MIUR COFIN2002 grant
``Stellar populations in the Local Group
as a tool to understand galaxy formation and evolution'' (P.I. M. Tosi).

\end{acknowledgements}

\bibliographystyle{aa}

\end{document}